\documentclass[a4paper,USenglish]{lipics-v2018}
\nolinenumbers
\usepackage{enumitem}
\usepackage{wrapfig}

\newtheorem{assumption}{\textbf{Assumption}}

\usepackage[vlined,ruled,linesnumbered,noresetcount]{algorithm2e}
\DontPrintSemicolon

\newcommand{\alref}[1]{\NlSty{\ref{#1}}}
\newcommand{\lref}[1]{line~\NlSty{\ref{#1}}}

\newcommand{\llref}[2]{lines~\NlSty{\ref{#1}}--\NlSty{\ref{#2}}}
\newcommand{\Llref}[2]{Lines~\NlSty{\ref{#1}}--\NlSty{\ref{#2}}}

\SetKw{Await}{await}
\SetKw{Goto}{goto}

\SetKw{CAS}{CAS}
\SetKw{DCAS}{DCAS}
\SetKw{FAS}{FAS}
\SetKw{FASAS}{FASAS}
\SetKw{Decide}{Decide}

\SetKw{KWTrue}{\small{true}}
\SetKw{KWFalse}{\small{false}}
\SetKwBlock{Loop}{loop}{end loop}
\SetKw{Int}{int}
\SetKw{Bool}{Boolean}
\SetKw{Array}{array}
\SetKw{Of}{of}
\SetKw{Const}{const}
\SetKw{Define}{Define}
\SetKw{Init}{init}
\SetKw{Goto}{goto}
\SetKw{Await}{await}
\SetKw{Break}{break}
\SetKw{True}{true}
\SetKw{False}{false}
\SetKw{Cont}{continue loop}
\SetKw{Local}{local}
\SetKw{Shared}{shared}
\SetKw{Null}{null}
\SetKw{Swap}{FAS}
\SetKw{CAS}{CAS}
\SetKw{Ref}{ref to}
\SetKwData{Side}{side}
\SetKwData{Left}{left}
\SetKwData{Right}{right}
\SetKwData{Other}{other}
\SetKw{Await}{await}
\SetKw{Goto}{goto}

\newif\ifcolourdraft
\ifcolourdraft

\else

\fi

\newcommand{\remove}[1]{}
\newcommand{\bc}[1]{\left\{ #1 \right\}}

\makeatletter
\newcommand{\removelatexerror}{\let\@latex@error\@gobble}
\makeatother

\title{Recoverable Consensus in Shared Memory} 

\author{Wojciech Golab}{University of Waterloo, Canada}{wgolab@uwaterloo.ca}{}{Author supported by the Natural Sciences and Engineering Research Council (NSERC) of Canada, and the Google Faculty Research Award program.}%mandatory, please use 

\Copyright{Wojciech Golab}
\authorrunning{Wojciech Golab}

\subjclass{Theory of computation~Concurrency}


\begin{document}
\maketitle

\begin{abstract}
Herlihy's consensus hierarchy is one of the most widely cited results in distributed computing theory.
It ranks the power of various synchronization primitives for solving consensus in a model
where asynchronous processes communicate through shared memory and fail by halting.
This paper revisits the consensus hierarchy in a model with crash-recovery failures,
	where the specification of consensus, called \emph{recoverable consensus} in this paper,
	is weakened by allowing non-terminating executions when a process fails infinitely often.
Two variations of this model are considered: independent failures, and simultaneous (i.e., system-wide) failures.
Universal primitives such as Compare-And-Swap solve consensus easily in both models,
	and so the contributions of the paper focus on lower levels of the hierarchy.
We make three contributions in that regard:
(i) We prove that any primitive at level two of Herlihy's hierarchy remains at level two if simultaneous
	crash-recovery failures are introduced.
This is accomplished by transforming (one instance of) any 2-process conventional consensus algorithm
	to a 2-process recoverable consensus algorithm.
(ii) For any $n > 1$ and $f > 0$, we show how to use $f+1$ instances of any
	conventional $n$-process consensus algorithm and $\Theta(f + n)$ read/write registers
	to solve $n$-process recoverable consensus when crash-recovery failures are independent,
	assuming that every execution contains at most $f$ such failures.
(iii) Next, we prove for any $f > 0$ that any 2-process recoverable consensus algorithm
	that uses TAS and read/writer registers requires at least $f+1$ TAS objects,
	assuming that crash-recovery failures are independent and every execution contains up to $f$
	such failures.
(iv) Lastly, we generalize and strengthen (iii) by proving that any universal construction of $n$-process recoverable consensus
	from a type $T$ with consensus number $n$ and read/write registers requires at least $f+1$ base objects of type $T$
	in executions with up to $f$ failures.

Results (ii) and (iii) establish a tight bound of $\Theta(f)$ on the space complexity
	of any 2-process recoverable consensus based on TAS for executions with
	$\Theta(f)$ independent failures.
This implies that 2-process recoverable consensus is not solvable in general using
	a finite number of TAS objects and read/write registers if failures are independent and their
	frequency is unbounded.
In contrast, (i) shows that the problem is solvable using only a single TAS object and $O(1)$ read/write
	registers when failures are simultaneous.
To our knowledge, this is the first separation between the two variations of the crash-recovery failure
	model under consideration with respect to the computability of consensus.
\end{abstract}

% !TEX root = main.tex
% !TEX spellcheck = en_US

\section{Introduction}
Herlihy's consensus hierarchy \cite{waitfree} ranks the power of various synchronization primitives for solving consensus -- a problem where processes agree on a decision chosen from a set of proposed values.
The position of a primitive $P$ in the hierarchy is defined by its \emph{consensus number}, which is the largest number $n$ such that
	$P$ used in conjunction with read/write registers solves $n$-process consensus in the asynchronous
	shared memory model.
%\footnote{The consensus number is infinite if no such $n$ exists.}
Consensus numbers answer the question of implementability in the following sense:
	if primitives $X$ and $Y$ have consensus numbers $n$ and $m$, respectively, where $n > m$,
	then $X$ can be used to implement $Y$ in a wait-free manner for up to $n$ processes,
	but $Y$ cannot be used to implement $X$ in a wait-free manner for more than $m$ processes.
%A primitive whose consensus number is greater than or equal to the maximum number of processes is called \emph{universal},
%	which means that it can be used to implement any other object in a wait-free manner.
Common synchronization primitives appear in Herlihy's consensus hierarchy at three distinct levels:
	read/write registers have consensus number 1, 
	so-called ``interfering'' primitives (e.g., Test-And-Set) as well as stacks and queues have consensus number 2,
	whereas Compare-And-Swap (CAS) has infinite consensus number.

The forthcoming adoption of non-volatile main memory (NVRAM) has revived interest in models of computation where processes may fail by crashing, and then recover.
Synchronization remains challenging in such models because the response of a shared memory
	operation is returned to a process using a volatile CPU register, which is part of the local state of
	process and is lost if a failure occurs before the response can be saved to non-volatile memory.
This observation calls into question the power of Read-Modify-Write (RMW) primitives for solving various synchronization tasks in the presence of crash-recovery failures, and exposes a new perspective
	for re-examining Herlihy's consensus hierarchy.

%The issue of computability in the crash-recovery model was studied earlier by Berryhill, Golab, and Tripunitara \cite{berry}, who proved that consensus remains sufficient to implement any shared
%	object type for any number of processes (i.e., remains universal).
%In particular, they showed that Herlihy's universal construction \cite{waitfree} satisfies a correctness criterion that generalizes Herlihy and Wing's widely-adopted linearizability property \cite{hw:linear} to the crash-recovery model.
%A straightforward proof shows also that CAS remains universal.
%On the other hand, little is known about how crash-recovery failures affect the synchronization
%	power of primitives that are not universal in Herlihy's hierarchy.
%	as it is much less clear how the introduction of crash-recovery failures affects their power
%	with respect to wait-free synchronization.

Building on the results of Berryhill, Golab, and Tripunitara \cite{berry}, who proved that consensus
	remains sufficiently powerful to implement any shared object type for any number of processes
	(i.e., remains universal) in the crash-recovery failure model,
	this paper advances the state of the art by establishing three fundamental results.
The contributions are phrased with respect to \emph{recoverable consensus} -- a natural weakening
	of consensus that accommodates executions with crash-recovery failures.
\begin{enumerate}
\item[(i)] We show how to use a single instance of any conventional 2-process consensus algorithm,
	and a constant number of read/write registers, to solve 2-process recoverable consensus 
	when failures are simultaneous (i.e., system-wide).
	Thus, any primitive at level two of Herlihy's hierarchy remains at level two when simultaneous
	crash-recovery failures are introduced.
\item[(ii)] For any $n > 1$ and $f > 0$, we show how to use $f+1$ instances of any
	conventional $n$-process consensus algorithm and $\Theta(f + n)$ read/write registers
	to solve $n$-process recoverable consensus when failures are independent,
	assuming that every execution has at most $f$ such failures.
\item[(iii)] Next, we prove for any $f > 0$ that any 2-process recoverable consensus algorithm
	that uses TAS and read/writer registers requires at least $f+1$ TAS objects,
	assuming that failures are independent and every execution contains at most $f$
	such failures.
\item[(iv)] Lastly, we generalize and strengthen (iii) by proving that any universal construction of $n$-process recoverable consensus
	from a type $T$ with consensus number $n$ and read/write registers requires at least $f+1$ base objects of type $T$
	in executions with up to $f$ failures.

%	Lastly, we show that there is no 2-process consensus algorithm 
%	that uses $f$ or fewer Test-And-Set (TAS) objects and any number of read/write registers
%	in the model with up to $f$ independent failures.
\end{enumerate}
Results (ii) and (iii) establish a tight bound of $\Theta(f)$ on the space complexity
	of any 2-process recoverable consensus based on TAS for executions with
	$\Theta(f)$ independent failures.
This implies that 2-process recoverable consensus is not solvable in general using
	a finite number of TAS objects and read/write registers if failures are independent.
In contrast, (i) shows that the problem is solvable using only a single TAS object and $O(1)$ read/write
	registers when failures are simultaneous.
To our knowledge, this is the first separation between the two variations of the crash-recovery failure
	model under consideration with respect to the computability of consensus.

% !TEX root = main.tex
% !TEX spellcheck = en_US
\newcommand{\IR}{\ensuremath{\textit{IR}}}
\newcommand{\inv}{\ensuremath{\textit{invoke}}}
\newcommand{\res}{\ensuremath{\textit{respond}}}
\newcommand{\trace}{\ensuremath{\textit{trace}}}

\section{Model} \label{sec:model}
Our model is based closely on Herlihy's \cite{waitfree}.

\textbf{Processes and objects.} \
The system comprises $n$ asynchronous processes, labeled $p_1, p_2, \ldots, p_n$,
	that communicate by accessing a finite number of typed shared objects.
The type of an object is described by a set of states, a distinguished initial state, a
	set of operations, and a state transition function that dictates the response 
	of a given operation applied in a given state.
Both processes and objects are deterministic.
Processes and objects can be modeled formally using I/O automata \cite{lynch}, but in this paper
	we describe their behavior less formally using pseudo-code.

\textbf{System and process states.} \
The \emph{state of the system} (\emph{state} for short) comprises the local state of each process
	(i.e., its program counter and local variables) as well as the state of each shared object
	(i.e., its current value).
There is a well-defined \emph{initial state of the system} in which the program counter of each process
	points to the beginning of its algorithm, its local variables hold their initial values,
	and each shared object is in the initial state prescribed by its type.
The state of the system changes in response to processes taking \emph{steps}, each of which entails
	bounded local computation followed by an atomic operation on one shared object
	(\emph{ordinary step}), or a crash-recovery failure (\emph{crash step}).
A crash step leaves the state of all shared objects unchanged, and resets the local state of one or more processes back to the initial state.
A process may recover after suffering a crash failure, in which case it resumes execution of its algorithm
	from the beginning.\footnote{A
process that recovers from a failure does not know the value of its own program counter immediately
	prior to failure, even if it uses a dedicated read/write register to record its own progress.
	This is because only one object or memory location can be accessed in one atomic step.}
Two variations of the crash failure are considered: an \emph{independent failure} affects a single process,
where as a \emph{simultaneous failure} affects all $n$ processes system-wide.
A crash step in the independent failure model identifies the failed process $p_i$ uniquely.

\textbf{Executions.} \
An \emph{execution} $E$ is a sequence of steps that is possible with respect to the
	algorithm prescribed for each process and the types of the shared objects, starting from
	the initial system state.
The set of such executions is prefix-closed.
%If execution $E$ if finite then the state of the system at the end of $E$ is well-defined.
A step $t$ is \emph{enabled} in state $s$ if $s$ is the state of the system at the end of some
	finite execution, and the sequence $G$ obtained by appending $s$ to $E$ (denoted $G = E \circ s$)
	is also an execution.
A crash step is enabled in any state.

\textbf{Consensus.} \
The algorithm executed by each process is a procedure that takes a \emph{proposal value} $v$
	as input, and returns a \emph{decision}.
The proposal value for each process $p_i$ is fixed for the duration of an execution,
	and is known to $p_i$ in the initial system state.
Different proposals map to different initial states.
In the classic model without failures (or, equivalently, with halting failures), the correctness properties
	of the algorithm are captured by the specification of the widely-studied \emph{consensus} problem:
\begin{enumerate}
\item {\emph{Agreement:}} distinct processes never output different decisions.
\item {\emph{Validity:}} each decision returned is the proposal value of some process.
\item {\emph{Wait-freedom:}} each process returns a decision after a finite number of its own steps.
\end{enumerate}
In this paper, we introduce a weakening of this problem called \emph{recoverable consensus}
	(or \emph{r-consensus} for short) that accommodates recovery from crash failures,
	which means that each process may attempt to compute a decision multiple times,
	as well as the inherent loss of liveness that may occur if a process fails repeatedly.
Specifically, the third correctness property is revised as follows:
\begin{enumerate}
\item[(3)] {\emph{Recoverable wait-freedom:}} each time a process executes its algorithm
	from the beginning, it either returns a decision after a finite number of its own steps,
	or crashes.
\end{enumerate}

Note that recoverable wait-freedom does not guarantee that the system as a whole makes progress in executions
where repeated failures prevent all processes from completing useful work.
This behavior is consistent with the definition of wait-freedom in the conventional model with permanent failures,
	where any $n$-process algorithm halts after $n$ crash failures.

\textbf{Consensus Numbers.} \
For any shared object type $T$, the \emph{consensus number} of $T$ with respect
	to a given failure model is the maximum
	number $m$ such that a finite number of objects of type $T$ and a finite number
	of additional read/write registers can be used to solve consensus (or recoverable
	consensus in our model) for $m$ processes, but not for $m+1$ processes.
If no such $m$ exists then the consensus number is infinite, and $T$ is called \emph{universal}.
%As we show later on in this paper, the consensus number of $T$ may depend on the failure assumption.

%\textbf{Valency.} \
%A state $s$ of the system at the end of a finite execution $E$ is called \emph{$v$-potent}
%	if there is some execution $E'$ that extends $E$ and in which some process decides $v$.
%State $s$ is called \emph{$v$-valent} if it is $v$-potent but not $v'$-potent for any $v' \neq v$.
%State $s$ is called \emph{univalent} if it is $v$-valent for some $v$.
%If $n = 2$ (i.e., the model has only two processes), then state $s$ is called \emph{bivalent} if it is not univalent.

% !TEX root = main.tex
% !TEX spellcheck = en_US

\newcommand{\twodef}[4]{\ensuremath{\left\{\begin{array}{rl}#1&\mbox{if\ \ }#2\\#3&\mbox{if\ \ }#4\end{array}\right.}}
\newcommand{\twodefo}[3]{\ensuremath{\left\{\begin{array}{rl}#1&\mbox{if\ \ }#2\\#3&\mbox{otherwise}\end{array}\right.}}

\section{Solving Recoverable Consensus Under Simultaneous Failures}\label{sec:upper1}
\begin{wrapfigure}{r}{0.5\textwidth}
%\vspace{-20pt}
%\begin{figure}[htbp]
\begin{flushleft}
\textbf{Shared variables:}
\begin{itemize}[leftmargin=5mm]
\item $P[1..2]$: \Array of proposal values, \Init $\bot$
\item $C$: conventional 2-process consensus
\item $D$: decision, \Init $\bot$
\end{itemize}
%\bigskip
\textbf{Private variables:}
\begin{itemize}[leftmargin=5mm]
\item $\Other$: process ID
\item $d$: decided value
\end{itemize}
%\bigskip
\end{flushleft}

%\removelatexerror
\begin{procedure}[H]
  \caption{() \protect \Decide\!($v$: proposal value) for process $p_i$, $i \in 1..2$}
  \leIf {$i = 1$} { $\Other := 2$ }{ $\Other := 1$ }
  \uIf {\emph{$P[i] = \bot$ $\wedge$ $P[\Other] = \bot$}}
  {  \nllabel{x:if}
  	$P[i] := v$\;		\nllabel{x:wP}
  	$d := C$.\Decide\!($v$)\;	\nllabel{x:C}
  	$D := d$\;		\nllabel{x:wD}
  	\Return $d$\;		\nllabel{x:retd}
  }
  \uElseIf {\emph{$D \neq \bot$}}
  { \nllabel{x:recD}
  	\Return $D$\;  \nllabel{x:recDret}
  }
  \uElseIf {\emph{$P[i] \neq \bot$ $\wedge$ $P[\Other] = \bot$}}
  {  \nllabel{x:inbotObot}
    	\Return $P[i]$\; \nllabel{x:inbotObotret}
  }
  \uElseIf {\emph{$P[i] = \bot$ $\wedge$ $P[\Other] \neq \bot$}}
  { \nllabel{x:ibotOnbot}
    	\Return $P[\Other]$\;  \nllabel{x:ibotOnbotret}
  }
  \uElse(// $P[i] \neq \bot \wedge P[\Other] \neq \bot$)
  { \nllabel{x:inbotOnbot}
  	\Return $P[1]$\; \nllabel{x:inbotOnbotret}
  } 
\end{procedure}
\caption{Transformation from 2-process conventional consensus to 2-process recoverable consensus.} \label{fig:trans}
%\end{figure}
%\vspace{-12pt}
\end{wrapfigure}
This section presents a technique for transforming any 2-process conventional consensus
	algorithm into a 2-process recoverable consensus algorithm that tolerates arbitrarily many
	simultaneous crash-recovery failures.
The transformation is presented in detail in Figure~\ref{fig:trans}.
It uses a shared array $P[1..2]$ to announce proposals, a conventional 2-process consensus algorithm
	$C$ to reach agreement in some scenarios (e.g., failure-free executions),
	and a shared variable $D$ to record the decision value computed using $C$.

Starting from the initial state where both elements of $P[1..2]$ are initialized to $\bot$,
	process $p_i$ records its proposal, executes $C$, and records the outcome in $D$ (\llref{x:if}{x:retd}).
Assuming that some process completes \lref{x:wD}, recovery from a failure is achieved easily
	by returning the value saved in $D$ (\llref{x:recD}{x:recDret}).
However, recovery from a failure prior to \lref{x:wD} is more difficult because if the failure
	occurred while some process $p_i$ was at \lref{x:C} then the algorithm must guarantee
	that $p_i$ does not access $C$ incorrectly (i.e., by resuming the interrupted execution of $C.\Decide$
	from the beginning) on recovery. % because $C$ is a conventional consensus algorithm.
Recovery in this scenario is accomplished by case analysis.
If a failure occurs before any process has completed \lref{x:wP} then the execution path on recovery
	is identical to the failure-free path since $P[1..2]$ and $C$ remain in their initial states.
On the other hand, if some process did complete \lref{x:wP} then the algorithm terminates after a
	bounded number of read operations, without updating shared memory.
If exactly one element of $P[1..2]$ is $\bot$ (\llref{x:inbotObot}{x:ibotOnbotret}) then
	the algorithm returns the unique recorded proposal.
Finally, if both elements of $P[1..2]$ are non-$\bot$ then the algorithm makes a deterministic
	choice among these two values and returns the chosen one (\llref{x:inbotOnbot}{x:inbotOnbotret}).
As presented, the algorithm chooses $P[1]$ but it could equally well chose $P[2]$, or
	the min or max of the two proposals.

The correctness properties of the algorithm are captured in Theorem~\ref{thm:x1sketch}.
Detailed analysis of the algorithm is deferred to Appendix~\ref{sec:upper1app} of \cite{arxiv} due to lack of space.

\begin{theorem}\label{thm:x1sketch}
The algorithm satisfies agreement, validity, and recoverable wait-freedom.
Furthermore, it has the same space complexity as the algorithm $C$.
\begin{proof}[Proof sketch]

\emph{Agreement}.
The decision is determined using $C$ at \lref{x:C} in executions where some
	process records this decision later on in $D$ at \lref{x:wD}.
If a failure occurs after one or both processes have written their proposals into $P[1..2]$ at \lref{x:wP}
	and before any process reaches \lref{x:wD}, then the decision is instead determined by inspecting the elements
	of $P$ on recovery at \llref{x:inbotObot}{x:inbotOnbotret} irrespective of the state of $C$.
\Llref{x:inbotObot}{x:ibotOnbotret} exploit the fact that $n = 2$, which allows
	a slower process (i.e., one who did not yet write its entry of $P[1..2]$)
	to adopt the proposal of a faster process (i.e., one who did write its entry of $P[1..2]$)
	and stop competing.

\emph{Validity}.
Only a proposed value can be written into $P[1..2]$ at \lref{x:wP}, swapped into $C$ at \lref{x:C},
	or written into $D$ at \lref{x:wD}.
Since the return value is always obtained from $D$ or from an element of $P$, validity follows.

\emph{Recoverable wait-freedom}.
The algorithm contains no loops, and the execution of $C$.\Decide at \lref{x:C} is wait-free.
A return statement is reached in any complete execution of the algorithm.
	
\emph{Space complexity}.
The algorithm uses three read-write registers in addition to $C$.
Therefore, its space complexity is equal asymptotically to the space complexity of $C$.
\end{proof}
\end{theorem}

% !TEX root = main.tex
% !TEX spellcheck = en_US

\section{Solving Recoverable Consensus Under Independent Failures}\label{sec:upper2}
This section presents a technique for transforming any $n$-process conventional consensus
	algorithm into an $n$-process recoverable consensus algorithm that tolerates
	up to a finite number $f$ of independent crash-recovery failures.\footnote{This parameter $f$ is the total number of failures incurred by all processes.}
The transformation is presented in detail in Figure~\ref{fig:trans2}, and uses $f+1$
	instances of the conventional consensus algorithm denoted by the array $C[0..f]$.
%This upper bound implies that any set of primitives that can solve consensus among $n$
%	processes in a wait-free manner in the conventional model with halting failures
%	can also solve consensus among $n$ processes in the crash-recovery model
%	with independent failures, using unbounded space.
To a first approximation, the transformation works by having each process $p_i$ access the 
	$f+1$ consensus algorithms in a for loop at \lref{xn:for} until the \Decide procedure
	is executed to completion without failing.
The array $R[1..n]$ is used at \lref{xn:if} and \lref{xn:inc} to determine which consensus algorithm in the array $C[0..f]$ will be accessed in the next iteration by each process,
	and hence to avoid unsafe access to these base objects.
Assuming that there are at most $f$ failures, this strategy ensures that $p_i$ eventually computes a decision
	because the total number of iterations required is at most one greater than the number of failures.
% TODO: cross-reference a lemma
The main technical challenge lies in ensuring agreement in cases when processes compute decisions
	in different iterations, using distinct instances of the conventional consensus algorithm.
This is accomplished by a pair mechanisms working in synergy.

In the first mechanism, a process $p_i$ that is executing iteration $k$ of the outer for loop checks at \llref{xn:forado}{xn:cado} whether
	a decision was reached in a lower-numbered iteration using $C[k']$ for some $k' < k$, and recorded in $D[k']$ at \lref{xn:wD},
	before $p_i$ proceeds to execute the consensus algorithm $C[k]$ at \lref{xn:C}.
If $D[k']$ holds such a decision value then $p_i$ adopts this value in lieu of its own proposal at \lref{xn:cado}.
This statement is inside the inner for loop and may be executed multiple times in one iteration of the outer for loop, in which case $p_i$ adopts the decision
	value corresponding to the largest possible $k'$.
This mechanism alone is not sufficient, however, since a race can occur between a process that is about to write $D[k-1]$
	and a process that is about to access $C[k]$.

\begin{wrapfigure}{r}{0.55\textwidth}
%\begin{figure}[htbp]
%\vspace{-12pt}
\begin{flushleft}
\textbf{Shared variables:}
\begin{itemize}[leftmargin=5mm]
\item $R[1..n]$: \Array of read/write register, \Init 0
\item $C[0..f]$: \Array of conventional wait-free $n$-process consensus objects
\item $D[0..f]$: \Array of read/write register, \Init $\bot$
\end{itemize}
%\bigskip
\textbf{Private variables:}
\begin{itemize}[leftmargin=5mm]
\item $k, k'$: integers, uninitialized
\item $d$: decision value, uninitialized
\end{itemize}
%\bigskip
\end{flushleft}

%\removelatexerror
\begin{procedure}[H]
  \caption{() \protect \Decide\!($v$: proposal value) for process $p_i$, $i \in 1..n$}
  \For {\emph{$k$ in $0..f$}} {  \nllabel{xn:for}
		\If {$R[i] = k$} {  \nllabel{xn:if}
			$R[i] := k + 1$\;   \nllabel{xn:inc}
			\tcp{check for a decision in a lower-numbered iteration}
			\For {\emph{$k' \in 0..(k-1)$}} {  \nllabel{xn:forado}
				\If {$D[k'] \neq \bot$} {  \nllabel{xn:cif}
					$v := D[k']$\;   \nllabel{xn:cado}
				}
			}
			$d := C[k]$.\Decide\!($v$)\;	\nllabel{xn:C}
			$D[k] := d$\;  \nllabel{xn:wD}
			\tcp{check for a collision with a higher-numbered iteration}
			\If {$k < f$} {  \nllabel{xn:ifp}
				\For {\emph{$z \in 1..n$, $z \neq i$}} {  \nllabel{xn:forp}
					\If {$R[z] > R[i]$} {  \nllabel{xn:ifpin}
						$d := \bot$\;  \nllabel{xn:ford}
					}
				}
			}
			\tcp{return decision if known }
			\If {$d \neq \bot$} {  \nllabel{xn:iffin}
				\Return $d$\;	\nllabel{xn:retd}
			}
		}
  }
\end{procedure}
\caption{Transformation from $n$-process conventional consensus to $n$-process recoverable consensus.} \label{fig:trans2}
% for executions with at most $f$ independent failures
%\end{figure}
\vspace{-12pt}
\end{wrapfigure}
The second mechanism deals with the above race condition at \llref{xn:ifp}{xn:ford} by inspecting the elements of array $R[1..n]$
	belonging to other processes.
If $p_i$ finds some element $R[z]$, $z \neq i$, holding an integer larger than $R[i]$ then $p_z$ is
	at least one iteration ahead of $p_i$.
In this case $p_i$ ``forgets'' the decision it computed earlier at \lref{xn:C} by resetting the private variable $d$ at \lref{xn:ford},
	and continues to the next iteration of the for loop;  the conditional statement at \lref{xn:iffin} bypasses the return statement at \lref{xn:retd}.
If $p_i$ does not find such a process $p_z$, then $p_i$ reaches \lref{xn:retd} where it returns the
	decision it computed in the current iteration of the outer for loop.
The two mechanisms combined ensure agreement (Lemma~\ref{lem2:agreement}) despite the fact that
	the consensus algorithms $C[0..f]$ may not all reach the same decision.

The correctness properties of the algorithm are captured in Theorem~\ref{thm:x2sketch}.
Detailed analysis of the algorithm is deferred to Appendix~\ref{sec:upper2app} of \cite{arxiv} due to lack of space.
The analysis is conditioned on the parameter $f$, which is the maximum number
	of failures possible on any execution.
The only part of the analysis relies on this parameter is the proof of recoverable wait-freedom.
% TODO: consider cutting down here
%Many parts of the proof refer to a process executing a specific iteration of the outer for loop, which
%	implicitly refers to the execution of \llref{xn:inc}{xn:retd} (i.e., the main body of the loop).
%It follows easily from the structure of the algorithm, particularly \llref{xn:if}{xn:inc},
%	that each process executes this part of the outer for loop at most once for any iteration number $k$,
%	even though it may execute \lref{xn:if} multiple times with the same $k$ due to failures.

\begin{theorem}\label{thm:x2sketch}
The algorithm satisfies agreement, validity, and recoverable wait-freedom in every execution with at most $f$ failures.
Furthermore, its space complexity is $O(fB + n)$ where $B$ denotes the space complexity of the $n$-process conventional consensus
	algorithm used to implement $C[0..f]$.
\break
\begin{proof}[Proof sketch]
\emph{Agreement}.
Suppose that $p_i$ and $p_j$ compute their return values.
If $p_i$ and $p_j$ do so using the same consensus algorithm $C[k]$ for some $k$ (i.e., in the same iteration),
	then the agreement property of $C[k]$ implies the agreement property of the recoverable consensus.
On the other hand, if $p_i$ and $p_j$ reach decisions using distinct consensus algorithms, say $C[k_i]$ and $C[k_j]$ for some $k_i < k_j$,
	then the mechanism at \llref{xn:ifp}{xn:ford} ensures that $p_i$ decides its response before $p_j$
	updates $R[j]$ in iteration $k_j$ at \lref{xn:inc}.
The mechanism at \llref{xn:forado}{xn:cado} then ensures that $p_j$ discovers $p_i$'s decision,
	and uses the decided value in place of its own proposal value when competing for $C[k_j]$.
This observation and the validity of $C[k_j]$ imply that $p_j$'s decision agrees with $p_i$'s.

\emph{Validity}.
A process only returns a value decided using some consensus algorithm $C[k]$ at \lref{x:C}.
The proposal value given as input to $C[k]$ is either the proposal value of the recoverable consensus algorithm
	for some process, or a value read from $D[k']$ at \lref{xn:cado} for some $k' < k$,
	which in turn is the decision reached in some lower-numbered iteration using $C[k']$.
A straightforward induction on the iteration number proves the validity of the recoverable consensus algorithm.
	
\emph{Recoverable wait-freedom}.
It follows from the structure of the algorithm, particularly \lref{xn:ifp}, that a process 
	reaches \lref{xn:retd} at the latest in its final iteration ($k = f$) and returns a response,
	unless it crashes in this final iteration.
We rule out the latter possibility by proving the following claim: at least $k$ failures must occur before any process
	begins iteration $k$.
The base case $k = 0$ follows trivially.
Suppose that the claim holds for iterations $k \in 0..x$ for some $x$, $0 \leq x < f$, and consider iteration $k = x + 1$.
Let $p_j$ be some process that reaches iteration $k = x + 1$, and consider how $p_j$ arrived in this scenario.
Without loss of generality, suppose that $p_j$ is the first process that reaches \lref{xn:if} with $k = x + 1$.

\noindent\emph{Case~1}: $p_j$ crashed in iteration $k = x$ after completing \lref{xn:inc}, and recovered with $R[j] = x + 1$.
Since $x$ failures have occurred by the induction hypothesis before any process reaches iteration $k = x$, it follows that
	$p_j$'s failure after \lref{xn:inc} is number $x+1$ or higher, as required.
	
\noindent\emph{Case~2}: $p_j$ completed iteration $k = x$ without crashing, and bypassed the return statement at \lref{xn:retd}.
Then $p_j$ executed \lref{xn:ford} in iteration $k = x$ after finding some other process $p_z$ such that $R[z] > R[j] = x$ by \lref{xn:ifpin}.
This implies that $p_z$ already completed \llref{xn:if}{xn:inc} in iteration $k = x + 1$ before $p_j$ started iteration $k = x + 1$,
	which contradicts the earlier assumption that $p_j$ is the first process that reaches \lref{xn:if} with $k = x + 1$ in the execution.

\emph{Space complexity}.
The algorithm uses $f+1$ instances of the $n$-process conventional consensus algorithm,
	and $O(f + n)$ additional read/write registers.
This implies the stated space complexity bound.
\end{proof}
\end{theorem}

% !TEX root = main.tex
% !TEX spellcheck = en_US

\section{Impossibility Result for Independent Failure Model}\label{sec:lower}
Any impossibility result for solving consensus in the conventional asynchronous model without failures applies also to
	solving recoverable consensus in the asynchronous model with crash-recovery failures.
This is because any execution that is possible in the conventional model is also admissible in the crash-recovery
	model, and because a violation of wait-freedom in such an execution implies a violation of recoverable wait-freedom
	(i.e., a process takes infinitely many steps and neither completes its algorithm nor fails).
Thus, the consensus number of an object with respect to recoverable consensus cannot exceed
	its consensus number in Herlihy's hierarchy \cite{waitfree}.
On the other hand, it is not known from prior results whether the power of a shared object type
	for solving consensus in a wait-free manner remains the same or is reduced once crash-recovery failures are introduced.
The question is answered partly in Section~\ref{sec:upper1} by showing that objects at level 2 in 
	Herlihy's hierarchy remain at level two if the model is relaxed to allow simultaneous crash-recovery failures,
	but the upper bound in Section~\ref{sec:upper2} does not imply the analogous result for independent failures
	because it assumes that the number of failures is bounded by a parameter $f$.
In particular, the transformation in Section~\ref{sec:upper2} uses $f+1$ instances of a conventional 2-process consensus algorithm
	to solve 2-process consensus with up to $f$ independent failures, and so its space complexity grows linearly with $f$.
In this section we show that this linear upper bound is tight asymptotically when the 2-process consensus algorithm
	is implemented using Test-And-Set (TAS).

\begin{wrapfigure}{r}{0.5\textwidth}
%\vspace{-12pt}
%\begin{figure}[htbp]
\begin{flushleft}
\textbf{Shared variables:}
\begin{itemize}[leftmargin=5mm]
\item $P[1..2]$: \Array of proposal values, \Init $\bot$
\item $C$: CAS object, \Init $\bot$
\end{itemize}
\end{flushleft}
%\removelatexerror
\begin{procedure}[H]
  \caption{() \protect \Decide\!($v$: proposal value) for process $p_i$, $i \in \bc{1, 2}$}
  \leIf {$i = 1$} { $\Other := 2$ }{ $\Other := 1$ }
  \uIf {\emph{$P[i] = \bot$ $\wedge$ $P[\Other] \neq \bot$}}
  { \nllabel{ex:if}
    	\Return $P[\Other]$\;  \nllabel{ex:retpo}
  }
  $P[i] := v$\;   \nllabel{ex:wP}
  \CAS\!($\&C, \bot, v$)\;	\nllabel{ex:CAS}
  \Return $C$\;		\nllabel{ex:retC}
\end{procedure}
\caption{A 2-process recoverable consensus algorithm demonstrating the possibility that a crash step may be a decision step.} \label{fig:cdexample}
%\end{figure}
%\vspace{-12pt}
\end{wrapfigure}

The proof technique used to establish the lower bound is an extension
	of the valency argument used by Herlihy \cite{waitfree} that accommodates crash-recovery failures.
A state $s$ of the system at the end of a finite execution $E$ is called \emph{$v$-potent}
	if there is some execution $E'$ that extends $E$ and in which some process decides $v$.
State $s$ is called \emph{$v$-valent} if it is $v$-potent but not $v'$-potent for any $v' \neq v$.
State $s$ is called \emph{univalent} if it is $v$-valent for some $v$.
If $n = 2$ (i.e., the model has only two processes), then state $s$ is called \emph{bivalent} if it is not univalent.

The conventional valency argument, when applied to the impossibility proof for read/write registers (Theorem~2 in \cite{waitfree}),
	constructs an execution at the end of which the state is bivalent, and any subsequent step by either process is a decision step.
This is accomplished by appending fragments to the execution starting from the initial state, where in each fragment one process
	takes steps without crashing until it is enabled to take a decision step.
In Herlihy's model, the construction must terminate eventually in some state $s$ since the algorithm is wait-free, and at that point both processes are about to take a decision step.
Moreover, the two decision steps necessarily lead to distinct decisions since $s$ is bivalent.
A contradiction is then reached by showing that either the two decision steps commute or one overwrites the effect of the other.

Herlihy's technique is not applicable directly in the crash-recovery failure model for two reasons.
First, the construction leading to state $s$ may never terminate since wait-freedom is relaxed (see Section~\ref{sec:model})
	to allow infinite executions in which processes fail infinitely often.
Second, even if the construction does terminate, two decision steps enabled in state $s$ may lead to the \emph{same} decision,
	which breaks the proof technique fundamentally.
As a simple example of this, consider the 2-process consensus protocol presented in Figure~\ref{fig:cdexample},
which uses similar principles to the transformation presented earlier in Section~\ref{sec:upper1}.
In this algorithm, processes record their proposal values in a shared array $P$ at \lref{ex:wP},
	and then use a CAS object $C$ at \lref{ex:CAS} to compute the decision, which is returned at \lref{ex:retC}.
If process $p_i$ crashes before completing \lref{ex:wP}, then on recovery it checks
	at \lref{ex:if} whether the other process $p_j$ already recorded its proposal.
If so, then $p_i$ yields to $p_j$ by returning $p_j$'s proposal at \lref{ex:retpo},
	and in this case $p_j$ wins $C$ uncontended as long as it takes sufficiently many steps without crashing.
Now consider an initially failure-free execution where processes $p_i$ and $p_j$ begin with distinct proposal values $v_i$ and $v_j$, respectively,
	and takes steps until a state $s$ is reached where each process is about to write $P$ at \lref{ex:wP}.
Then $s$ is bivalent because either process can still win $C$ in some extension of the execution.
If $p_i$ takes an additional step by writing its proposal to $P[i]$ at \lref{ex:wP} then the new state $s'$ is also bivalent for the same reason as $s$.
If $p_i$ takes yet another step from state $s'$ by executing CAS at \lref{ex:CAS} then the execution
	becomes $v_i$-valent, however the same holds if $p_j$ transitions out of $s'$ by crashing since on
	recovery it must adopt $p_i$'s proposal as $P[i] \neq \bot$ and $P[j] = \bot$.
There are also two transitions out of $s'$ to another bivalent state, namely if
	$p_i$ crashes instead of executing \lref{ex:CAS}, or if $p_j$ completes \lref{ex:wP}.

We adapt Herlihy's proof technique to the crash-recovery failure model by resolving the above technicalities
	through a simplifying assumption that in fact strengthens our impossibility result:

\begin{assumption}\label{as1}
	For any execution, a process $p_i$ takes a crash step if and only if:
	\begin{enumerate}
	\item $p_i$'s previous step was its first access in this execution to some TAS object; and
	\item $p_i$ is the lowest-numbered process that participates in the execution
	\end{enumerate}
\end{assumption}
%In the context of algorithms that use Test-And-Set and read/write registers, this principle is applied as follows.
%First, a crash failure always occurs after the first access to a given Test-And-Set object by a given process, and in no other situation.
Assumption~\ref{as1} ensures that an execution involving two processes and $m$ TAS objects contains at most $m$ failures.
Furthermore, for each state $s$ and each process $p_i$, there is at most one step that $p_i$ is enabled to execute from $s$.
%Second, we assume that the crash failure and its preceding TAS are executed atomically (in that order), and represent a single state transition.
%This implies that at most two transitions are possible out of a given state.
With the above in mind, the impossibility result is captured in Theorem~\ref{thm:tas}.

% What is a TAS object?

\begin{theorem}\label{thm:tas}
For any integer $f > 0$, there is no algorithm that uses at most $f$ readable Test-And-Set (rTAS) objects and any number of read/write registers,
	and solves 2-process recoverable consensus in the crash-recovery model with independent failures,
	even if there are at most $f$ such failures.
\begin{proof}
Without loss of generality, assume the restricted failure model where all executions obey Assumption~\ref{as1},
	and define valency with respect to the subset of executions allowable in this restricted model.
Fix $f > 0$, and let $p_1, p_2$ be the two processes.
Suppose for contradiction that a 2-process recoverable consensus algorithm $A$ does exist using at most $f$ rTAS objects
	and some finite number of read/write registers.
%Recall that all executions in the restricted model contain at most $2f$ failures since there are only two processes.

Starting in the initial system state, construct an execution fragment where $p_1$ takes steps solo until it is enabled to execute a decision step.
Append a second fragment where $p_2$ takes steps solo until it is also about to execute a decision step.
Continue appending fragments of steps by $p_1$ and $p_2$ in an alternating sequence until a bivalent state $s$ is reached where both processes are enabled
	to execute decision steps.
The construction must succeed eventually by the wait-freedom of $A$ since every execution contains at most $f$ failures under Assumption~\ref{as1}.
Furthermore, since there are exactly two transitions out of the bivalent state $s$, the two decision steps necessarily lead to distinct decisions
	as in Herlihy's proof \cite{waitfree}.
Suppose that $p_1$'s step leads to a $v_1$-valent state, and $p_2$'s step leads to a $v_2$-valent state, where $v_1 \neq v_2$.

\emph{Case~A:} both processes are poised to apply operations on distinct objects.
Then these decision steps commute, meaning that they lead to the same state irrespective of the order in which they are applied.
If $p_1$ takes its step first followed by $p_2$, then this leads to a $v_1$-valent state $s'$, and so there is some
	execution fragment $F$ where both process continue to take steps from $s'$ and return $v_1$.
If $p_2$ takes its step first followed by $p_1$, then this leads to a $v_2$-valent state $s''$, which is indistinguishable
	to both processes from $s'$.
Thus, both processes are enabled to execute the sequence of steps in $F$ from $s''$ as well and return $v_1$,
	which contradicts the observation that $s''$ is $v_2$-valent.

\emph{Case~B:} both processes are poised to apply TAS operations on an object $T$, and neither has accessed $T$ in the execution leading to state $s$.
Let $s'$ be the state obtained from $s$ by allowing $p_1$ to apply its TAS and crash.
Then $s'$ is $v_1$-valent and so there is some sequence of steps in $F$ from $s'$ where $p_1$ returns $v_1$.
Let $s''$ be the $v_2$-valent state obtained from $s$ by allowing $p_2$ to apply its TAS, then allowing $p_1$ to apply its TAS and crash.
Then $s''$ is indistinguishable to $p_1$ from $s'$ and so $p_1$ is enabled to execute the sequence of steps in $F$ from $s''$ as well and return $v_1$,
	which contradicts the observation that $s''$ is $v_2$-valent.

\emph{Case~C:} both processes are poised to apply TAS operations on an object $T$, and at least one process, say $p_1$, has accessed $T$ already in the execution leading to state $s$.
Then $T$ holds the value 1 in state $s$, and so both TAS operations return 1 and leave the value of $T$ unchanged.
Thus, the two decision steps commute, and a contradiction is reached by the same argument as in Case~A.

\emph{Case~D:} both processes are poised to access the same base object and at least one process is poised to read, or at least one process is poised to crash,
	or both processes are poised to write.
We will show that the following claim holds (or its analog with $p_1$ and $p_2$ interchanged):
	\begin{quote}
	The system state $s'$ obtained by allowing $p_1$ and $p_2$ to take their decision steps from $s$, in that order,
	is indistinguishable to $p_2$ from the state $s''$ obtained by allowing $p_2$ alone to take its decision step from $s$.
	\end{quote}
The claim implies that $s'$ is $v_1$-valent, and so there exists an execution fragment $F$ in which $p_2$ takes steps solo from $s'$ and terminates with a $v_1$ return value.
Similarly, the claim implies that $s''$ is $v_2$-valent, and yet $p_2$ is enabled to execute the same sequence $F$ of steps from $s''$ as from $s'$,
		since $s''$ is indistinguishable to $p_2$ from $s'$.
Thus, $p_2$ can terminate with a $v_1$ return value in some extension of the execution ending in $s''$, which contradicts $s''$ being $v_2$-valent.
To complete the proof of Case~D, it remains to show that the claim holds for the constructed bivalent state $s$.

\emph{Subcase~D-1:} at least one process, say $p_1$, is about to read or crash.
Then $p_1$'s decision step has no effect on shared memory, and so the claim holds.
	
\emph{Subcase~D-2:} both processes are poised to write the same read/write register.
Then $p_2$'s decision step overwrites the effect of $p_1$'s decision step, and so the the claim holds.
\end{proof}
\end{theorem}

%.\footnote{We reach a similar observation with respect to simultaneous failures by analyzing carefully the pseudo-code in Figure~\ref{fig:trans}.}

\begin{corollary}\label{cor:tas}
Any 2-process recoverable consensus algorithm in the crash-recovery model with independent failures 
	that uses readable Test-And-Set (rTAS) objects and any number of read/write registers
	requires more than $f$ rTAS objects in executions with up to $f$ failures. 
\end{corollary}

The $\Omega(f)$ lower bound on space complexity implied by Corollary~\ref{cor:tas} is tight with
	respect to the upper bound presented in Section~\ref{sec:upper2}.
In particular, if 2-process conventional consensus is implemented using a single TAS (or rTAS) object and no additional read/write registers,
	then the transformation from Section~\ref{sec:upper2} yields a 2-process recoverable consensus algorithm with $O(f)$ space complexity.
%\footnote{The 2-process recoverable consensus algorithm uses $O(f)$ TAS objects and $O(f)$ read/write registers.}

% !TEX root = main.tex
% !TEX spellcheck = en_US

\section{Universal Primitives}\label{sec:upper3}

Herlihy \cite{waitfree} discusses several universal primitives for consensus.
%	compare-and-swap, memory-to-memory move and swap, augmented queue, fetch-and-cons, and sticky byte.
The universality of Compare-And-Swap for recoverable consensus is established by the same algorithm
	as for consensus: a memory word is initialized to a special value $\bot$, and every process attempts to 
	swap its own input in place of this value.
Exactly one process succeeds, and its input becomes the decided value.
Assuming that all inputs are different from $\bot$, the algorithm ensures that processes agree on a valid decision
	regardless of how many times they attempt to swap in their input, and regardless of whether failures are simultaneous or independent.
It is also straightforward to prove the universality of several other primitives with respect to recoverable consensus:
	a queue augmented with a \emph{peek} operation that returns but does not remove the first item;
	a fetch-and-cons object, which atomically threads an item onto the front of a linked list;
	as well as memory-to-memory move and swap.
It suffices to modify the constructions described in \cite{waitfree} slightly to avoid repeated applications of the universal
	primitive to the same memory word by a process recovering from a failure,
	for example by scanning the linked list before attempting to insert a new node using fetch-and-cons.

The observation that universal primitives in Herlihy's hierarchy tend to remain universal in the crash-recovery model
	leads to the following a natural question: is there a general construction of recoverable consensus from consensus.
The transformation presented in Section~\ref{sec:upper2} answers this question only partially because its
	space complexity grows linearly with both $n$ and the bound $f$ on the number of failures in an execution.
As a result, a practical instantiation of this technique can only tolerate a bounded number of failures.
Another approach, suggested by an anonymous referee, is to use a single $\infty$-consensus base object and $n$ single-writer registers
	as follows: ``when accessing the object the $k$'th time (recorded in its single-writer register),
	process $i$ can simply pretend that it is process $kn+i$ and access the object accordingly.''
While this construction produces correct outputs, it has two drawbacks related to the unbounded growth of the process IDs.
First, the $\infty$-consensus object may internally use process IDs to index array elements, as in the construction
	from memory-to-memory move and swap \cite{waitfree}.
Even if this is not an issue, additional registers are used record (a lower bound on) the number of failures experienced by individual processes,
	and hence their size grows as $\Theta(\log_2 f)$.
Thus, the space complexity of the suggested construction is unbounded unless $f$ is fixed ahead of time.

We conclude this section by proving formally that bounded space is inherently impossible in a generic
	construction of recoverable consensus from consensus and read/write registers.
Genericity in this context is defined formally as follows:
\begin{definition}[Genericity]\label{defg}
A construction of $n$-process recoverable consensus from $n$-process consensus and read/write registers
	is \emph{generic} if the following property holds in all executions:
	if a process $p_i$ accesses an $n$-process consensus base object $C$ and then crashes,
	then $p_i$ does not access $C$ again in the remainder of the execution.
\end{definition}
Definition~\ref{defg} captures the point that the correctness of $C$ is contingent on the absence of failures,
	and hence it is potentially unsafe for a process to access $C$ both before and after a crash failure.
As an example, the constructions presented earlier in Sections~\ref{sec:upper1} and \ref{sec:upper2} both
	satisfy Definition~\ref{defg}.

\begin{theorem}\label{thm:univimp}
For any integer $f > 0$, there is no generic algorithm that uses at most $f$ $n$-consensus objects and any number of read/write registers,
	and solves $n$-process recoverable consensus in the crash-recovery model with independent failures,
	even if there are at most $f$ such failures and at most two processes participate.
\begin{proof}
Suppose for contradiction that an $n$-process recoverable consensus algorithm $A$ does exist using at most $f$ $n$-consensus objects
	and some finite number of read/write registers.
The analysis is a generalization of the proof of Theorem~\ref{thm:tas} in Section~\ref{sec:lower}.
We consider the subset of executions where only two processes participate, say $p_1$ and $p_2$, and
	where the following assumption (analogous to Assumption~\ref{as1}) holds.
\begin{assumption}\label{as2}
	For any execution, a process $p_i$ takes a crash step if and only if:
	\begin{enumerate}
	\item $p_i$'s previous step was its first access in this execution to some $n$-process consensus object; and
	\item $p_i$ is the lowest-numbered process that participates in the execution
	\end{enumerate}
\end{assumption}
Valency is defined with respect to this restricted set of executions, and we construct an execution that leads to a state $s$
	where each process $p_i$ is enabled to take a decision step that leads to a $v_i$-valent state.
Any access to an $n$-process consensus base object in an execution is represented by a single atomic step.

\emph{Case~A:} both processes are poised to apply operations on distinct objects.
The analysis is the same as Case~A in the proof of Theorem~\ref{thm:tas}.

\emph{Case~B:} both processes are poised to apply \Decide operations on a consensus object $C$, and neither has accessed $C$ in the execution leading to state $s$.
%The analysis is the same as Case~B in the proof of Theorem~\ref{thm:tas}, with $T$ replaced by $C$ and TAS replaced by \Decide.
Since we assume that $p_1$ and $p_2$ are the only processes participating in the execution, the base object $C$ is in its initial state in $s$.
Let $s'$ be the state obtained from $s$ by allowing $p_1$ to apply its \Decide, which sets the decision of $C$ to $v_1$, and crash.
Then $s'$ is $v_1$-valent and so there is some sequence of steps in $F$ from $s'$ where $p_1$ returns $v_1$.
Let $s''$ be the $v_2$-valent state obtained from $s$ by allowing $p_2$ to apply its \Decide first, which sets the decision of $C$ to $v_2$,
	then allowing $p_1$ to apply its \Decide and crash.
Then $s''$ is indistinguishable to $p_1$ from $s'$ despite $C$ reaching different decisions in these two states.
This is because the genericity of $A$ (Definition~\ref{defg}) ensures that $p_1$ never accesses $C$ again after its decision step under consideration.
Thus, $p_1$ is enabled to execute the sequence of steps in $F$ from $s''$ as well and return $v_1$,
	which contradicts the observation that $s''$ is $v_2$-valent.

\emph{Case~C:} both processes are poised to apply \Decide operations on a consensus object $C$, and at least one process, say $p_1$, has accessed $C$ already in the execution leading to state $s$.
Then $C$ has already settled to some decision value $v$ prior to state $s$.
As a result, both \Decide operations applied from state $s$ return $v$ and leave the decision of $C$ unchanged.
Thus, the two decision steps commute, and a contradiction is reached by the same argument as in Case~A.

\emph{Case~D:} both processes are poised to access the same base object and at least one process is poised to read, or at least one process is poised to crash,
	or both processes are poised to write.
The analysis is the same as Case~D in the proof of Theorem~\ref{thm:tas}.
\end{proof}
\end{theorem}

\begin{corollary}\label{cor:univimp}
Any generic $n$-process recoverable consensus algorithm in the crash-recovery model with independent failures 
	that uses $n$-process consensus objects and any number of read/write registers requires more than $f$ 
	$n$-process consensus objects in executions with up to $f$ failures. 
\end{corollary}

Theorem~\ref{thm:univimp} and Corollary~\ref{cor:univimp} imply a lower bound of $\Omega(f)$ on the space complexity
	of generic constructions for $n$-process recoverable consensus from $n$-process consensus, as well as from
	from any type with consensus number $n$.
These results hold independently of $n$.

% Can we prove that 

%Section 5 starts by assuming that the results using crash-stop are "stricter" than crash/recovery, therefore the findings are more general. In principle I agree with this, but in practice, are you sure that all your assumptions (e.g., recoverable wait-freedom) do not hamper this inclusion? It is maybe worth to specify that in the paper in order to avoid leaving readers with open questions.
% !TEX root = main.tex
% !TEX spellcheck = en_US

\section{Related Work} \label{sec:related}

Herlihy's seminal paper on wait-free synchronization \cite{waitfree} defined the consensus
	hierarchy, which ranks the synchronization power of any shared object type in terms of 
	the maximum number of processes for which it can solve consensus.
Some types, such as Compare-And-Swap, are universal (infinite consensus number),
	meaning that when used in conjunction with read/write registers, they can solve consensus
	or implement any other object type in a wait-free manner for arbitrarily many processes.
Read/write registers, on the other hand, are at the bottom of the hierarchy with consensus number one,
	meaning that they cannot solve consensus for even two processes.
Level two of the hierarchy features a set of commonly implemented objects
	including Test-And-Set, Swap, Fetch-And-Add, as well as stacks and queues.
Herlihy's hierarchy builds on several earlier lower bounds, including Loui and Abu-Amara's
	proof that consensus cannot be solved using Test-And-Set and read/write registers
	for three or more processes \cite{loui1987memory}.

Herlihy's hierarchy treats different synchronization primitives as distinct shared objects, and assumes
	in defining the consensus number of a given type that that one object of this type is used in
	conjunction with any number of read/write registers.
Jayanti proved that in some cases using multiple objects of the same type with consensus number $k$ (e.g., ``weak-sticky'') makes it possible to solve consensus for $k+1$ processes \cite{jayarobust}.
This observation violates the principle of robustness, which states that multiple objects with consensus number at most $k$ cannot solve consensus for more than $k$ processes.
Ellen et al.\ further showed that when the operations of some widely-supported types
	at level 2 in Herlihy's hierarchy (e.g., fetch-and-add and test-and-set) are composed into a single more powerful type, then
	consensus can be solved for any number of processes \cite{ellenhier}.
Since Herlihy's computability-based hierarchy collapses under this alternative (and more realistic) notion of composition,
	Ellen et al.\ proposed that the power of synchronization primitives for solving consensus
	should instead by ranked in terms of space complexity \cite{ellenhier}.
This alternative hierarchy is based on obstruction-free consensus, and places readable test-and-set 
	at the same level as read/write registers, similarly to our results for wait-free consensus
	under independent crash-recovery failures (Section~\ref{sec:lower}). 
%In comparison, our upper and lower bound constructions adopt a variation of Herlihy's model and
%	both computability and space complexity.

The set of object types that are implementable in a wait-free manner for any
	number of processes using objects with consensus number two is called \emph{Common2}
	\cite{common2}.
Every type at level two of Herlihy's consensus hierarchy is in Common2, and Common2
	is known to include many of the types at level two, including Test-And-Set,
	Swap, and stacks \cite{common2, common2x}.
It is not known whether queues belong to Common2, and intense research effort
	has been devoted to answering this question \cite{david04, barnes93, ellen12, fatourou11}.
	
% 2, 4, 5, 6, 8, 14

Herlihy \cite{waitfree} proved the universality of Compare-And-Swap by exhibiting a construction that
	uses a bounded
	number of consensus objects and read/write registers to simulate a linearizable~\cite{hw:linear}
	implementation of any object type given its sequential specification.
Theoretical and practical aspects of such universal constructions in the standard asynchronous
	model with halting failures have been widely studied
	\cite{arelock,fatourou11,ellen05,kogan12,timnat14,jayanti98}.
Berryhill, Golab, and Tripunitara extended the universality result for CAS to the crash-recovery
	model with independent failures by showing that Herlihy's universal construction remains correct in this model \cite{berry}.
This result proves that wait-free synchronization remains possible even when the failure model
	is relaxed, but does not answer fundamental questions regarding the relative power
	of different shared object types under variations of the crash-recovery model.

Correctness properties for shared objects that tolerate crash-recovery failures are proposed in several papers.
Contributions in this space are extensions of the widely-adopted linearizability property of Herlihy and Wing~\cite{hw:linear}.
Aguilera and Fr{\o}lund proposed \emph{strict linearizability} \cite{sven}, which requires operations interrupted by a failure
	to take effect either before the failure or not at all.
Guerraoui and Levy proposed a relaxed condition called \emph{persistent atomicity} in the context of message passing systems \cite{robem},
	which allows an operation the take effect even after a failure, before the next operation invocation of the same process.
Berryhill, Golab, and Tripunitara defined \emph{recoverable linearizability} \cite{berry}, which builds on Guerraoui and Levy's
	definition by restoring locality -- the desirable property that an execution involving multiple objects is correct if and only if
	its projection onto each individual object is correct.
Izraelevitz, Mendes and Scott recently proposed \emph{durable linearizability} \cite{ims16} in a model with simultaneous failures,
	hardware buffering, and relaxed consistency.
Durable linearizability is equivalent to both persistent atomicity and recoverable linearizability
	in a model where crash failures are simultaneous \cite{ims16}.

Herlihy's universal construction \cite{waitfree} uses a ``helping mechanism'' that ensures
	progress by allowing an operation invoked by a slower process to take effect by the
	action of another faster process.
Censor-Hillel, Petrank and Timnat formalized helping and analyzed its necessity \cite{help}.
Attiya, Castaneda and Hendler defined alternative notions of helping, and used them
	to separate queues and stacks in terms of the possibility of wait-free implementation
	from Common2 objects \cite{help2}.
%Their results lend evidence to the hypothesis that queues may be inherently more difficult
%	to implement that stacks from objects with consensus number two.

The crash-recovery failure model features prominently in recent work on recoverable
	mutual exclusion \cite{gr:rme,gh:rme2,jaya1}.
For the class of algorithms that use commonly supported single-word synchronization
	primitives, there exists a gap in terms of time complexity upper bounds between conventional
	mutual exclusion with halting failures and recoverable mutual exclusion
	with independent crash-recovery failures.
This observation suggests, but does not prove, that crash-recovery failures may reduce
	the power of read-modify-write primitives for efficient synchronization.
% !TEX root = main.tex
% !TEX spellcheck = en_US

\section{Conclusion} \label{sec:conclusion}
This paper introduced possibility and impossibility results for solving consensus
	in the asynchronous model with crash-recovery failures.
For any primitives at level two of Herlihy's consensus hierarchy \cite{waitfree}
	(e.g., Test-And-Set), the transformation presented in Section~\ref{sec:upper1} shows that
	the power of this primitive for solving consensus is unchanged if the standard
	asynchronous model is weakened by the introduction of simultaneous crash-recovery failures.
The combined results of Sections~\ref{sec:upper2} and \ref{sec:lower}
	prove that the same observation does not hold if the	model is weakened further
	by the introduction of independent crash-recovery failures.
This is because	
	Test-And-Set can no longer be used in conjunction with read/write registers
	to solve consensus among two processes in bounded space.
These results separate the model with independent crash-recovery failures 
	both from the model with simultaneous crash-recovery failures, and 
	the standard model with halting failures, in terms of the computability of consensus.
%A broader investigation of how crash-recovery failures affect the power of other
%	primitives for wait-free synchronization is left as future work.

\bibliographystyle{plainurl}
\bibliography{main}

\newpage
\appendix
% !TEX root = main.tex
% !TEX spellcheck = en_US

\section{Analysis for Section~\ref{sec:upper1}}\label{sec:upper1app}

\begin{lemma}\label{lem:xeti}
In any execution, the consensus algorithm $C$ is accessed correctly by each process $p_i$.
\begin{proof}
We must show that each process accesses $C$ at most once despite the possibility of crash-recover failures.
This follows from the condition at \lref{x:if}, the fact that $p_i$ assigns a non-$\bot$ value
	to $P[i]$ at \lref{x:wP} before accessing $C$, and the fact that $C$ is only accessed at \lref{x:C}.
\end{proof}
\end{lemma}

\begin{lemma}\label{lem:agr1}
In any execution, if the decision value computed using $C$ is recorded in $D$ at \lref{x:wD}
	then no process returns a different decision value.
\begin{proof}
First, note that by the specification of $C$ and by Lemma~\ref{lem:xeti},
	processes agree on the decision value of $C$, and so the value written to $D$ at \lref{x:wD}
	is fixed even if multiple processes execute this step.
Furthermore, this value is the proposal of the first process to complete \lref{x:C}.
Let $p_j$ be this process.
Then in any execution of the \Decide procedure by $p_j$, either there is crash failure before $p_j$ returns
	a value, or $p_j$ returns at \lref{x:retd} or \lref{x:recDret}.
In both cases $p_j$ returns its own proposal.
Now consider the value returned by some execution of \Decide of the other process, $p_i$.
We will show that this value is also $p_j$'s proposal, as required for agreement.

\emph{Case~1:} $p_i$ returns at \lref{x:retd}.
	Then $p_i$ returns $p_j$'s proposal, which it computes at \lref{x:C}.

\emph{Case~2:} $p_i$ returns at \lref{x:recDret}.
	Then $p_i$ returns the non-$\bot$ value stored in $D$, which is the decision
	value computed using $C$ at \lref{x:C}, which is $p_j$'s proposal.

\emph{Case~3:} $p_i$ returns at \lref{x:inbotObotret}.	
	Then $P[i] \neq \bot$ and $P[j] = \bot$ hold when $p_i$ reads $P[j]$ at \lref{x:inbotObot}.
	This implies that a simultaneous failure of both processes occurred after $p_i$ last executed \lref{x:wP},
	as otherwise $p_i$ would have completed \llref{x:C}{x:retd} and never reached \lref{x:inbotObotret}.
	Furthermore, such a failure occurred before $p_j$ ever completed \llref{x:wP}{x:wD} because $p_i$
	reads $P[j] = \bot$ at \lref{x:inbotObotret}, and so any subsequent execution of
	\lref{x:if} by $p_j$ causes $p_j$ to read $P[i] \neq \bot$
	at \lref{x:if} and branch to \lref{x:recD}.
	This contradicts the earlier assumption that $p_j$ reaches \lref{x:wD} in the same execution.

\emph{Case~4:} $p_j$ returns at \lref{x:ibotOnbotret}.
	Then $p_i$ reads $p_j$'s proposal from $P[j]$ at \lref{x:ibotOnbotret},
	and then returns this value.

\emph{Case~5:} $p_j$ returns at \lref{x:inbotOnbotret}.
	Then $P[i] \neq \bot$ and $P[j] \neq \bot$ hold when $p_i$ reads $P[j]$ at \lref{x:inbotObot}.
	As in Case~3, this implies that a failure occurred after $p_i$ last executed \lref{x:wP}.
	Furthermore, such a failure occurred before $p_j$ completed \llref{x:wP}{x:wD} as otherwise
	$p_i$ would have read $D \neq \bot$ earlier at \lref{x:recD}, and returned
	at \lref{x:recDret} instead of \lref{x:inbotOnbotret}.
	This leads to a contradiction, as in Case~3, because any subsequent execution of
	\lref{x:if} by $p_j$ causes $p_j$ to read $P[i] \neq \bot$	at \lref{x:if},
	and then bypass \lref{x:wD}.
\end{proof}
\end{lemma}

\begin{lemma}\label{lem:agr2}
In any execution, if \lref{x:wD} is never completed by any process then no two processes return different values.
\begin{proof}
Suppose that \lref{x:wD} is never completed.
Then all executions of \Decide that produce a response terminate at \lref{x:inbotObotret}, \lref{x:ibotOnbotret}, or \lref{x:inbotOnbotret}.

\emph{Case~1:} neither processes completes \lref{x:wP}.
Then $P$ remains in its initial state throughput the execution,
	every process proceeds from \lref{x:if} to \lref{x:wP},
	and no process returns a response since we assume that \lref{x:wD} is never reached.
(This is possible in a finite execution where neither process has taken sufficiently
many steps to compute a decision.)

\emph{Case~2:} only one process, say $p_j$, completes \lref{x:wP}.
Then a response can only be returned at \lref{x:inbotObotret} or \lref{x:ibotOnbotret},
	and this response is $p_j$'s proposal read from $P[j]$.
Thus, agreement holds.

\emph{Case~3:} both processes complete \lref{x:wP}.
Then both processes first complete \lref{x:if} while $P$ is in its initial state,
	and then complete \lref{x:wP} before the next failure (if any) occurs.
Since we assume that \lref{x:wD} is never completed in this execution, it follows
	that both processes subsequently either remain at \llref{x:C}{x:wD} where no
	response is returned,
	or fail simultaneously and then return responses only at \lref{x:inbotOnbotret}.
Agreement then follows from the deterministic choice of proposal value at \lref{x:inbotOnbotret}.
This value can be different from the one returned in Case~2, however that does not break the algorithm
	since Case~2 and Case~3 are disjoint.
\end{proof}
\end{lemma}

\begin{theorem}\label{thm:x}
The algorithm satisfies agreement, validity, and recoverable wait-freedom.
Furthermore, it has the same space complexity as the algorithm $C$.
\begin{proof}

\emph{Agreement}.
The decision returned by the algorithm is computed either using $C$, or by 
	inspecting the proposal values in $P[1..2]$.
If a decision is computed using $C$ and then recorded in $D$ at \lref{x:wD}
	then agreement follows from Lemma~\ref{lem:agr1}.
Otherwise, it follows from Lemma~\ref{lem:agr2}.

\emph{Validity}.
The algorithm only returns a value read from $P$ or $D$.
Any value written to $P$ is the proposal value of some process.
Any value written to $D$ is the decision computed using $C$,
	which is the proposal value of some process because $C$
	ensures validity by its specification and by Lemma~\ref{lem:xeti}.

\emph{Recoverable wait-freedom}.
The algorithm contains no loops, and the execution of $C$.\Decide at \lref{x:C} is wait-free
	by the specification of $C$ and by Lemma~\ref{lem:xeti}.
	
\emph{Space complexity}.
The algorithm uses three read-write registers in addition to $C$.
Therefore, its space complexity is equal asymptotically to the space complexity of $C$.
\end{proof}
\end{theorem}

% !TEX root = main.tex
% !TEX spellcheck = en_US

\section{Analysis for Section~\ref{sec:upper2}}\label{sec:upper2app}
The analysis is conditioned on the parameter $f$, which is the maximum number
	of failures possible on any execution.
The only part of the analysis that assumes at most $f$ failures it the proof of recoverable wait-freedom
(Lemma~\ref{lem2:waitfree}).
Many parts of the proof refer to a process executing a specific iteration of the outer for loop, which
	implicitly refers to the execution of \llref{xn:inc}{xn:retd} (i.e., the main body of the loop).
It follows easily from the structure of the algorithm, particularly \llref{xn:if}{xn:inc},
	that each process executes this part of the outer for loop at most once for any iteration number $k$,
	even though it may execute \lref{xn:if} multiple times with the same $k$ due to failures.

\begin{lemma}\label{lem2:xeti}
For any $k \in 0..f$, the consensus object $C[k]$ is accessed correctly by each process $p_i$
	in any execution.
\begin{proof}
We must show that each process accesses $C[k]$ at most once despite the possibility of crash-recover failures.
This follows from lines \alref{xn:if} and \alref{xn:inc}, which precede the access to $C[k]$ at \lref{xn:C},
	as well as from the initialization of $R[1..n]$ to 0.
Specifically, the order of lines \alref{xn:inc} and \alref{xn:C} ensures that $R[i] > k$ by the time $p_i$ accesses $C[k]$,
	and the monotonic growth of $R[i]$ ensures that $p_i$ never accesses $C[k]$ again irrespective of failures
	due to the condition at \lref{xn:if}.
\end{proof}
\end{lemma}

\begin{lemma}\label{lem2:validity}
The algorithm satisfies validity in every execution.
\begin{proof}
First, note that any value returned by the algorithm at \lref{xn:retd} is the decision computed
	using one of the consensus algorithms $C[0..f]$ at \lref{xn:C}.
Since Lemma~\ref{lem2:xeti} ensures the validity of these decisions, it suffices to prove
	the following claim:
	\begin{quote}
	\emph{Every value $v$ used at \lref{xn:C} is the proposal value of some process
	executing the recoverable consensus algorithm.}
	\end{quote}
We will prove this claim by induction in the iteration number $k$.
In the base case $k = 0$, \lref{xn:cado} is bypassed due to the condition $k > 0$ at \lref{xn:cif},
	and so process $p_i$ can only access $C[k]$ at \lref{xn:C} using its own proposal value, as required.
Now suppose that the claim holds for iterations $0..k$, and consider iteration $k+1$ if it exists (i.e., if $k < f$).
If \lref{xn:cado} is bypassed in iteration $k+1$ then the claim follows as in the base case.
Otherwise $p_i$ observes $D[k'] \neq \bot$ at \lref{xn:cif} for some $k' < k$, which implies that some
	process reached a decision using $C[k']$ and then recorded it in $D[k']$ at \lref{xn:wD}.
Since $D[k'] \neq \bot$ is a stable property, and since $C[k']$ ensures validity by Lemma~\ref{lem2:xeti},
	it follows that the value written to $D[k']$ and then read by $p_i$ at \lref{xn:cado} is the proposal value
	used by some process during its own iteration $k' < k$.
By the induction hypothesis, this value is the proposal of some process executing the recoverable consensus algorithm, as required.
\end{proof}
\end{lemma}

\begin{lemma}\label{lem2:agreement}
The algorithm satisfies agreement in every execution.
\begin{proof}
Suppose that distinct processes $p_i$ and $p_j$ return values $d_i$ and $d_j$, respectively, in some execution.
	
\emph{Case~A:} $p_i$ and $p_j$ terminate in the same iteration $k$ of the outer for loop, $0 \leq k \leq f$.
Then both processes return the decision computed using $C[k]$ at \lref{xn:C}.
Since Lemma~\ref{lem2:xeti} ensures that this consensus algorithm ensures agreement,
	it follows that $d_i = d_j$, as required.

\emph{Case~B:} $p_i$ and $p_j$ terminate in distinct iterations $k_i$ and $k_j$, respectively.
Without loss of generality, suppose that $0 \leq k_i < k_j \leq f$.
Since $p_i$ returns in iteration $k_i$, it follows that $p_i$ executed \llref{xn:ifp}{xn:ford}
	in this iteration after computing the decision using $C[k_i]$ at \lref{xn:C},
	and did not find any other process $p_z$ such that $R[z] > R[i]$ (\lref{xn:ifpin}).
Thus, at the time when $p_i$ reached \lref{xn:ford} in iteration $k_i$,
	no process had reached \lref{xn:inc} in its own iteration $k_i$ or higher.
As a result, any process $p_z$ that executes the for loop at \llref{xn:forado}{xn:cado} in iteration $k_i + 1$ or higher
	discovers $p_i$'s return value $d_i$ in $D[k_i]$.
This holds even if multiple processes write $D[k_i]$ at \lref{xn:wD} because such decisions
	are computed using $C[k_i]$, whose agreement property follows from Lemma~\ref{lem2:xeti}.
To complete the proof, we use induction on the iteration number to prove the following claim:
\begin{quote}
	\emph{If $p_z$ executes \Decide at \lref{xn:C} in iteration $k_i + 1$ or higher then it
	uses a proposal value $v$ equal to $p_i$'s return value, $d_i$.}
\end{quote}
The claim implies agreement between $p_i$ and $p_j$ because the validity property of $C[(k_i + 1)..f]$,
	which follows from Lemma~\ref{lem2:xeti}, ensures that $p_z$ can only return $d_i$ in the iteration under consideration.

In the base case, iteration $k_i + 1$, process $p_z$ observes $D[k_i] \neq \bot$ at
	\lref{xn:cif}, and moreover $D[k_i] = d_i$, as noted earlier in this proof.
Since $k_i$ is the largest value of $k'$ considered in the for loop at \llref{xn:forado}{xn:cado},
	it follows that $p_z$ then proceeds to access $C[k_i + 1]$ at \lref{xn:C} with a proposal value $v = d_i$, as required.
Next, suppose that the claim holds for iterations $k_i + 1$ up to $k_i + m$ for some $m \geq 1$, and 
	consider iteration $k_i + m + 1$, if it exists.
In this iteration, $p_z$ observes $D[x] \neq \bot$ at \lref{xn:cif} for at least one value of $x$, namely $x = k_i$.
Let $y$ be the largest array index for which $D[y] \neq \bot$ is observed,
	where $y < k_i + m + 1$ by the upper limit of the for loop at \lref{xn:forado}.
Then the value $D[y]$ read by $p_z$ is the decision recorded at \lref{xn:wD} by some process that 
	accessed $C[y]$ at \lref{xn:C} in iteration $y$ of that process.
Since $C[y]$ ensures validity by Lemma~\ref{lem2:xeti}, it follows from the induction hypothesis
	that this value is $d_i$, as required.
\end{proof}
\end{lemma}

\begin{lemma}\label{lem2:waitfree}
The algorithm satisfies recoverable wait-freedom in every execution with at most $f$ failures.
\begin{proof}
Since Lemma~\ref{lem2:xeti} ensures the wait-freedom of the consensus objects $C[0..f]$, it follows
	that each iteration of the outer for loop is wait-free.
Therefore, it suffices to show that each process $p_i$ either reaches \lref{xn:retd} after a finite number
	of iterations and returns a response, or crashes.\footnote{If this does property does not hold then $p_i$ could complete iterations $0..f$ and
	reach the end of the pseudo-code in Figure~\ref{fig:trans2} without returning a response.}
If $p_i$ does not terminate or crash in the first $f$ iterations (i.e., $0 \leq k < f$) then
	it reaches another iteration with $k = f$.
If it does not crash in this iteration then it computes a decision value $d \neq \bot$ at \lref{xn:C} by the validity of $C[k]$,
	which follows from Lemma~\ref{lem2:xeti}.
It then bypasses \llref{xn:ifp}{xn:ford} due to the condition $k < f$ tested at \lref{xn:ifp},
	and reaches \llref{xn:iffin}{xn:retd} with $d \neq \bot$.

To complete the proof, it remains to show that if $p_i$ does not return a decision in the first $f$ iterations
	then it eventually executes \llref{xn:inc}{xn:retd} in the last iteration where $k = f$.
This amounts to proving that $p_i$ cannot fail during this final iteration, or in other words
	that $f$ failures have already occurred by the time $p_i$ reaches $k = f$.
To that end, we will prove the following claim by induction:
\begin{quote}
	For any $k$ such that $0 \leq k \leq f$, at least $k$ failures must occur before any process reaches $k$.
\end{quote}
The base case, $k = 0$, follows trivially.
Next, suppose that the claim holds for $k$ up to and including some $x$, $0 \leq x < f$, and consider the claim for $k = x + 1$.
By the induction hypothesis, $x$ failures have occurred before any process reaches iteration $k = x$.
Let $p_j$ be some process that reaches iteration $k = x + 1$, and consider how $p_j$ arrived in this scenario.
Without loss of generality, suppose that $p_j$ is the first process that reaches \lref{xn:if} with $k = x + 1$.

\emph{Case~1}: $p_j$ crashed in iteration $k = x$ after completing \lref{xn:inc}, and recovered with $R[j] = x + 1$.
Since $x$ failures have occurred by the induction hypothesis before any process reaches $k = x$, it follows that
	$p_j$'s failure after \lref{xn:inc} is number $x+1$ or higher, as required.
	
\emph{Case~2}: $p_j$ completed iteration $k = x$ without crashing, and bypassed the return statement at \lref{xn:retd}.
Then $p_j$ executed \lref{xn:ford} in iteration $k = x$ after finding some other process $p_z$ such that $R[z] > R[j] = x$ by \lref{xn:ifpin}.
This implies that $p_z$ already completed \llref{xn:if}{xn:inc} in iteration $k = x + 1$ before $p_j$ started iteration $k = x + 1$,
	which contradicts the earlier assumption that $p_j$ is the first process that reaches \lref{xn:if} with $k = x + 1$ in the execution.
\end{proof}
\end{lemma}

\begin{theorem}\label{thm:x2}
The algorithm satisfies agreement, validity, and recoverable wait-freedom in every execution with at most $f$ failures.
Furthermore, its space complexity is $O(fB + n)$ where $B$ denotes the space complexity of the $n$-process conventional consensus
	algorithm used to implement $C[0..f]$.
\begin{proof}
Agreement, validity, and recoverable wait-freedom are established in Lemmas \ref{lem2:agreement}, \ref{lem2:validity} and \ref{lem2:waitfree}.
The space complexity follows from the use of $f+1$ instances of the $n$-process conventional consensus algorithm,
	and $O(f + n)$ additional read/write registers.
\end{proof}
\end{theorem}

\end{document}